\documentclass[%
 reprint,
 floatfix,
superscriptaddress,
nofootinbib,
 amsmath,amssymb,
 aps,
prd,
showkeys,
]{revtex4-2}
\usepackage{booktabs}
\usepackage{centernot}
\usepackage{graphicx}
\usepackage{dcolumn}
\usepackage[dvipsnames]{xcolor}
\usepackage{bm}
\usepackage{hyperref}
\usepackage[mathlines]{lineno}
\usepackage{xcolor}
\usepackage{amsmath}
\usepackage[capitalize]{cleveref}
\usepackage[normalem]{ulem}
\usepackage{csquotes}

\usepackage{enumitem}

\newcommand*\mnras{MNRAS}
\newcommand*\aap{A\&A}

\newcommand*\jcap{J. Cosmology Astropart. Phys.}
\newcommand*\physrep{Phys.~Rep.}

\newtheorem{statement}{Statement}

\def\d{{\rm d}}

\begin{document}

\title{Incompatibility of fine-structure constant variations at recombination with local observations}

\author{L\'eo Vacher}
\email{lvacher@sissa.it}
\affiliation{International School for Advanced Studies (SISSA), Via Bonomea 265, 34136, Trieste, Italy}
\author{Nils Sch\"oneberg}\thanks{Léo Vacher and Nils  Sch\"oneberg contributed equally to this work and are to be considered identically as first authors.}
\affiliation{Institut de Ciències del Cosmos (ICCUB), Facultat de F\'isica, Universitat de Barcelona (IEEC-UB), Mart\'i i Franqués, 1, E08028 Barcelona, Spain}
\date{\today}

\begin{abstract}
Some attempts of easing the critical Hubble tension present in modern cosmology have resorted to using variations of fundamental constants, such as the fine-structure constant, at the time of recombination. In this article we demonstrate that there are critical hurdles to construct such viable models using scalar fields, due to the striking precision of local constraints on the fine-structure constant stability. These hurdles demonstrate that in single-field models one has to extremely fine-tune the shape of the potential and/or the initial conditions. Indeed, for single field models in a potential that is not fine-tuned we can put a generic bound at recombination of $\Delta \alpha/\alpha_0 < 5\cdot 10^{-4}$ (95\% CL).
\end{abstract}

\maketitle


\section{Introduction}
\enlargethispage*{1\baselineskip}

Thanks to the incredible experimental and theoretical effort undertaken within the last decades, the precision of the measurement and understanding of the underlying cosmological model has steadily increased. However, this increased precision has uncovered new tensions between different measurements of cosmological parameters -- such as the Hubble constant (which specifies the current expansion rate) -- within the cosmological standard model ($\Lambda$CDM, involving a cosmological constant and cold collisionless dark matter). This Hubble tension between local distance ladder measurements using Cepheids and determinations from the Planck satellite measuring the cosmic microwave background (CMB) anisotropies have now reached a significance of beyond $5\sigma$ \cite{Riess2022,Planck2018,Verde2023}.

It has been recently proposed that an early variation of the fundamental constants 
could ease 
this Hubble tension
by delaying the time at which recombination occurs \cite{Planckalphame2016,Hart:2021kad,H0olympics,Lee2023,Chluba2023}. Such primordial variations can be constrained directly using the CMB anisotropies and spectral distortions, but these constraints remain loose enough to allow for the existence of significant deviations with respect to the local values measured on Earth \cite{Hart:2017ndk,Hart2023,Tohfa2023,Chluba2023}. However, to remain consistent from a high energy physics perspective, a space-time dependent fundamental constant -- as the fine-structure constant $\alpha$, quantifying the intensity of the electromagnetic force -- must be induced by a fundamental field implemented at the Lagrangian level (for a review see for example \cite{Uzan2011}). Such fields are for example unavoidable in string theory, with the scalar dilaton field, partner mode of the graviton, inducing a variation of all the standard model's gauge couplings \cite{Damour1994}. 

In general, any model that shows a displacement of such a coupled field causes a variation of the fine-structure constant in the early Universe. However, as we will detail within this work, if a relatively large relative variation of the fine-structure constant is desired, this displacement should occur as early as possible. As such, investigating scalar fields that become dynamical around the time of recombination are of particular interest. The prime example of such fields is an axion-like particle (ALP) with a decay constant such that it becomes dynamical around recombination. Indeed, such a field has been used to supply an era of early dark energy contribution \cite{axion-cosmology,axion-dark-energy}, which has also been shown to provide a successful path in order to ease the Hubble tension \cite{Poulin2018,Poulin2019,Poulin2023}, though not without caveats \cite{Hill2020,Ivanov2020,D'Amico2021,Murgia2021,Amon2022,Goldstein2023,Gsponer2023}. These ALP are deeply motivated both from high energy, as solution to the strong CP problem and modes from string theory \cite{Peccei2008,Svrcek2006}.
Moreover, such ALP could produce some parity violating signal in the cosmic microwave background which claims to have been detected in Planck data \cite{Minami2020,Diego-Palazuelos2022,Diego-Palazuelos2023}. Note, however, that this claim of detection has been contested by recent analyses as \cite{Zagatti2024}. As such, an important question to address is whether such an ALP could be also coupled to electromagnetism and induce a variation of the fine-structure constant, further alleviating the $H_0$ tension (and/or overcoming the shortcomings of EDE). As proposed in \cite{Flambaum2009}, such a field fully coupled to electromagnetism through both a scalar and a pseudo-scalar term is expected to be present in general scalar-tensor theories.

We will see that such a scenario would be in conflict with the extremely tight constraints imposed by laboratory data on Earth today. 
Along with our exploration of the Swampland conjectures in \cite{Schoneberg2023}, we will 
derive an \enquote{almost no-go} theorem, stating that a simple early varying fine-structure constant scalar field model cannot possibly provide fine-structure constant variations large enough to be cosmologically relevant without an extreme level of fine-tuning in either the potential or the field initial conditions.
\newpage

We will start by presenting the theoretical background for varying fine-structure constant models in \ref{ssec:theory}, followed by the derivation of an \enquote{almost no-go theorem} for models with early fine-structure constant variation in \cref{ssec:nogo}. The validity of this theorem will then be demonstrated in two models: axion like particles coupled to electromagnetism in \cref{ssec:ede} and a toy model with an hyperbolic tangent potential in \cref{ssec:nogo}. Finally, we will discuss the extent and the limits of our study followed by our conclusions in \cref{sec:conclusion}.

\section{Theoretical motivation} \label{sec:theory}

In order to remain consistent with the most basic principles of physics, any variation of the fundamental constants of nature must be implemented at the Lagrangian level, promoting the constants to scalar fields. In \cref{ssec:theory}, we will present the formalism allowing for a consistent variation of the fine-structure constant. From this theoretical background, we will show in \cref{ssec:nogo} using simple derivations that such scalar field models can not allow for an early variation of the fine-structure constant and while remaining compatible with local data without an extraordinary amount of fine tuning.   

\subsection{Fields coupled to electromagnetism}
\label{ssec:theory}
The fine-structure constant $\alpha$ is a dimensionless gauge coupling quantifying the intensity of the electromagnetic force. It represents a favorable observable in order to investigate the possible variation of the fundamental constants, as its impact on physics is well known and it is possible to measure its value with a great accuracy throughout cosmic history using multiple independent probes \cite{Martins2022}.\footnote{Additionally, $\alpha$ is dimensionless, and only the stability of dimensionless constant can be investigated unambiguously. For a discussion see for example \cite{Duff2002}.} As far as we know, the only self-consistent path to promote $\alpha$ as a dynamical quantity, is to introduce new fields at the Lagrangian level which are responsible for its variation. If $\alpha \to \alpha(\phi)$ then the electromagnetic kinetic Lagrangian must be modified as
\begin{equation}
    -\frac{1}{4}F_{\mu\nu}F^{\mu\nu}\to -\frac{1}{4}B_F(\phi)F_{\mu\nu}F^{\mu\nu}
\end{equation}
in order to preserve the U(1) gauge invariance of the theory, such that the fine-structure constant evolves as
\begin{equation}
\frac{\Delta \alpha}{\alpha_0}= \frac{\alpha\left(\phi\right)-\alpha_0}{\alpha_0}= B_F(\phi)^{-1} -1~,
\end{equation}
where $\alpha_0\sim 1/137$ is the value of the fine-structure constant measured in laboratory.\footnote{This formalism for fine-structure constant variations was presented in \cite{Olive2002} as a generalization of the models of Bekenstein \cite{bekensteinOriginal} and Sandvik et al. \cite{Sandvik2002}, which are the limiting cases where the scalar field is proportional to the electron charge (leading through a change of variables to $\alpha(\phi) = \alpha_0e^{2\phi}$and $B_F(\phi)\propto e^{-2\phi}$ for a canonically normalized field $\phi$).} Including a kinetic and potential term for the scalar field, its full Lagrangian can be written as 
\begin{align}
\mathcal{L}= & -\frac{1}{2}\partial_\mu\phi\partial^\mu\phi - V(\phi) \nonumber\\
&-  \frac{1}{4}B_F(\phi)F_{\mu\nu}F^{\mu \nu}+...
\end{align}

Knowing that the possible fine-structure constant variation allowed by experiments are extremely restricted, the scalar coupling can be typically linearized as 
\begin{equation}
    B_F(\phi) \simeq 1 + \zeta (\phi-\phi_0)
\end{equation}
where $\zeta= \partial_\phi B_F|_{\rm \phi=\phi_0}$ (with the sign convention of \cite{Olive2002}). This approximation appears to accurately model the time variation of the fine-structure constant through cosmic history for a wide range of models. While we will use this linearization in the remainder of the main text, we also discuss how generic this approach is in \cref{app:linearization}. 

From the expression of $B_F$, one can express the variations of the fine-structure constant as
\begin{equation}\label{eq:delta_alpha}
    \frac{\Delta\alpha}{\alpha_0} \simeq -\zeta(\phi-\phi_0) ~.
\end{equation}
The amplitude of variations of such a scalar field through cosmological times, and hence the allowed variations of $\alpha$, are sharply restricted by the atomic clocks measurement of \citep{Filzinger2023} providing a bound of
\begin{equation}\label{eq:atomicclocks}
\frac{1}{\alpha_0}\frac{\d\alpha}{\d t}\Bigg|_{z=0}=
(1.8\pm2.5)\cdot 10^{-19}/{\rm yr}~.
\end{equation}
As such, in the absence of screening mechanisms, any significant variations of the field from the CMB responsible for a different fine-structure constant at recombination must brutally rapidly decrease to match local constraints.

In addition, direct astrophysical measurements of $\Delta \alpha/\alpha_0$ itself (which are made using quasars) lead to $\Delta\alpha/\alpha_0(1\leq z\leq2.5)\sim10^{-6}$ (see \cref{it:fine-structure}). For comparison, these measurements can be converted to give an approximate estimation of the fine structure constant drift rate of $(\mathrm{d}\alpha/\mathrm{d}t)/\alpha_0~|_{z\simeq 1.5}\sim10^{-16} h/\mathrm{yr}$. Such estimation can be done by comparing the measurement of quasars at different redshifts (here we used a polynomial regression of the data up to various orders). Hence, local data on $(\mathrm{d}{\alpha}/\mathrm{d}t)/\alpha_0$ are more stringent than local data by around two and a half orders of magnitude.

\subsection{Almost a no-go theorem}
\label{ssec:nogo}

Let us now investigate further how the experimental bounds can constrain the presence of varying $\alpha$ during recombination. For this, let us look at the two ingredients that are present for such a model: 
\begin{enumerate}
    \item A relevantly large $\Delta \alpha/\alpha_0$ at the time of recombination (by definition)
    \item A small $\d \ln \alpha/\d \ln a$ today in order to avoid atomic clock constraints
\end{enumerate}
As we are going to see, in a standard formalism of coupling based on a single scalar field, these two conditions are highly incompatible. 

First, we must consider what we mean by relevantly large $\Delta \alpha/\alpha_0$ during recombination. For our purpose, we will choose a benchmark value of $\Delta \alpha/\alpha_0(z_\mathrm{cmb}) = 10^{-2}$, which would in principle completely resolve the Hubble tension (this value was computed with a simple code such that it leads to the same angular sound horizon while raising $H_0$ to $73.04 \mathrm{km/s/Mpc}$, the value preferred by \cite{Riess2022}). 

However, it is already clear that such high values of $\alpha(z_\mathrm{cmb})$ are not compatible with the Planck 2018 angular power spectra measurements, since it is possible to infer a constraint of \cite{Hart:2017ndk} 
\begin{equation}
    \frac{\Delta\alpha}{\alpha_0}\left(z_\mathrm{cmb}\right)= (-0.7 \pm 2.5)\cdot 10^{-3}~.
\end{equation}
However, we note that this constraint is dependent on the geometrical degeneracies present in the CMB and as such might be eased when introducing additional model parameters such as curvature, variation of the neutrino mass, dark energy equation of state, etc. Instead, the bound we derive below will be mostly independent of the specifics of the evolution of the various cosmic energy densities and their perturbations. We also note that it has been derived assuming typically a constant shift of the fine structure constant. Finally, we will see that the bound we derive is even tighter than that obtained from the CMB itself.


Now, let us introduce the two quantities
\begin{align}
    \mathfrak{D}\equiv  & \frac{\Delta \alpha}{\alpha_0}(z_{\rm cmb})  = - \zeta \Delta \phi = - \zeta \int_{\ln a_\mathrm{cmb}}^{0} \frac{\d \phi}{\d \ln a} \d \ln a,\\
    \epsilon \equiv & \left.\frac{\d \ln \alpha}{\d \ln a }\right|_{z=0} = - \zeta \left.\frac{\d \phi}{\d \ln a}\right|_{z=0}.
\end{align}
In this notation we can succinctly summarize our constraints as $\epsilon \simeq (1.76 \pm 2.44) \cdot 10^{-9}/h$ from \cref{eq:atomicclocks}, where we use the usual definition of $h = H_0/[100\mathrm{km/s/Mpc}]$. Comparing with a benchmark value of $\mathfrak{D} \simeq 10^{-2}$ gives us a ratio of $\mathfrak{D}/\epsilon \sim 10^{7}$.
If we re-parameterize the field speed evolution through some function $U(z)$ such that
\begin{equation}
    \frac{\d \phi}{\d \ln a} = U(z) \left.\frac{\d \phi}{\d \ln a}\right|_{z=0}.
\end{equation}
we can immediately relate the two expressions as
\begin{equation}
    \mathfrak{D}= \epsilon \int_{0}^{\ln 1+z_\mathrm{cmb}} U(z) \d \ln(1+z).
\end{equation}
Importantly, this relation is \emph{independent} of the precise value of $\zeta$, since it simultaneously rescales  $\mathfrak{D}$ and $\epsilon$\,. As such, the combination of a given desired $\mathfrak{D}$ at early times and a given constraint on $\epsilon$ from atomic clocks observations gives a coupling-independent constraint on the necessary evolution of the field speed.

Looking at the rough order of magnitude of this constraint, the problem becomes quickly apparent. In only roughly 7 $e$-folds (from recombination to now), the relative field speed $U(z)$ has to vary by approximately a factor of $\mathfrak{D}/\epsilon \sim 10^{7}$. It turns out that with mild assumptions, this combination is almost impossible. Below, we will put a bound on $\mathfrak{D}/\epsilon$ from simple considerations and translate it into a bound on $\mathfrak{D}$\,.
To show this, let us slightly rewrite the equations of motion for the scalar field speed\footnote{We neglect the possible couplings of the field to the other sectors of the Universe, expected to be present in such models in the Klein-Gordon and the Friedmann equations through terms of the form $\zeta\Delta\phi\rho_i$\cite{Sandvik2002}. Given the allowed values for $\zeta$, the impact on the field evolution is expected to be largely subdominant, see also \cref{sec:conclusion}.}
\begin{equation}\label{eq:kg_eom}
    \frac{\d \dot{\phi}}{\d \ln a} + 3 \dot{\phi} = - \frac{1}{H}\frac{\d V}{\d \phi} 
\end{equation}
and explicitly integrate to obtain
\begin{equation}\label{eq:kg_sol}
    \dot{\phi} = C a^{-3} - a^{-3}\int \frac{a^3}{H} \frac{\d V}{\d \phi} \d \ln a.
\end{equation}
Let us, for a moment, focus on the homogeneous equation. We are going to come back to the inhomogeneous part in a bit. The homogeneous part $\dot{\phi} = Ca^{-3}$ tells us that from Hubble drag the field speed can only decay at a rate of $a^{-3}=(1+z)^3$. This immediately places a tight constraint on $U(z) = \dot{\phi}/H \cdot H_0/\dot{\phi}_0$. Indeed, since $v(0)=1$ by definition, we would have $U(z) = (1+z)^3 \cdot H_0/H$ in this case. It is now easy to put a conservative bound on the integral as
\begin{widetext}
\begin{align}\label{eq:analytical}
    \mathfrak{D}/\epsilon & \approx \int_0^{\ln 1+z_\Lambda} (1+z)^3 \frac{H_0}{H} \d \ln (1+z) + \frac{1}{\sqrt{\Omega_m}}\int_{\ln 1+z_\Lambda}^{\ln 1+z_\mathrm{cmb}} (1+z)^3 (1+z)^{-3/2} \d \ln (1+z)\nonumber\\ 
    & =  \int_0^{\ln 1+z_\Lambda} (1+z)^3 (1+z)^{-\frac{3}{2}(1+w)}\d \ln (1+z)  + \frac{2}{3 \sqrt{\Omega_m}} \left[(1+z_\mathrm{cmb})^{3/2}-(1+z_\Lambda)^{3/2}\right] \nonumber \\
    & < \int_0^{\ln 1+z_\Lambda} (1+z)^3 (1+z)^{-\frac{3}{2}(1+(-1))}\d \ln (1+z)  + \frac{2}{3 \sqrt{\Omega_m}} \left[(1+z_\mathrm{cmb})^{3/2}-0\right] \\
    & = \frac{1}{3} \left[(1+z_\Lambda)^3-1\right]  + \frac{2}{3 \sqrt{\Omega_m}} \left[(1+z_\mathrm{cmb})^{3/2}\right] \nonumber\\
    & < \frac{1}{3} \left[(1+10)^3-1\right]  + \frac{2}{3 \cdot 0.3} \left[(1+1200)^{3/2}\right] \approx 9.3 \cdot 10^4 \nonumber
\end{align}
\end{widetext}

If we require our benchmark value of $\mathfrak{D}/\epsilon \sim 10^7$, then this is obviously a contradiction. Indeed, we can use this bound to impose limits on $\mathfrak{D} = \Delta \alpha(z_\mathrm{cmb})/\alpha_0$ as we show below. However, we first aim to give a few details on the calculation.

In the first line of \cref{eq:analytical} we have split the integral into one part where we assume matter domination to hold (above some $z_\Lambda$), and one part where the dark energy might be important. 

In the second line we have combined the first and second Friedmann equations
\begin{equation}
    \frac{2(\dot{H}+H^2)}{H^2} = - \frac{\frac{8\pi G}{3}(\rho+3P)}{\frac{8\pi G}{3}\rho} = -(1+3 w)~,
\end{equation}
with $w = P/\rho$ being the total equation of state in the Universe, which can quickly be simplified to
\begin{equation}
    \frac{\d\ln H}{\d \ln a} = \frac{\dot{H}}{H^2} = -\frac{1+3w}{2}-1 = -\frac{3}{2}(1+w)~,
\end{equation}
which gives the relation of $H/H_0 = (1+z)^{3/2 (1+w)}$ used here. Note that $w$ at this point can and does depend on time, and is simply defined as $w(z) = P(z)/\rho(z)$. Now, for any normal contents of the Universe, we have $w \in [-1,1]$ and this allows us to quickly bound the first summand in the third line of \cref{eq:analytical}.\footnote{In principle a $w<-1$ could break this argument, but this is practically irrelevant for two reasons: i) the first summand turns out to have a negligble 0.5\% contribution to the total sum, and for it to have a 50\% contribution one would need $w \simeq -2.4$, which is strongly excluded by any late-time data and ii) the bound assumed for $z_\Lambda$ of around $10$ is very conservative, and even if a non-trivial dark energy model with $w \ll -1$ would be used, also in this case the bounds could be made tighter by choosing a less conservative~$z_\Lambda$\,.} Let us stress again that this derivation is intended to propose a conservative bound rather than an accurate estimate. In the same direction, we also bounded $1+z_\Lambda>0$ for the second summand. In the fourth line we simply evaluate the remaining integral. Finally, we bound $z_\Lambda <10$ and $z_\mathrm{cmb}<1200$ as well as $\Omega_m > 0.1$ (to get $\sqrt{\Omega_m} > 0.3$)  in the fifth line to give the final upper bound. Note that these are very conservative upper bounds, since CMB data typically require a smaller recombination redshift and measurments of baryonic acoustic oscillations (BAO) or supernovae of type Ia (SNIa) require $z_\Lambda \sim 0.33 \ll 10$ and $\Omega_m \sim 0.3$, see below. 

Since the computation is always dominated by the matter-dominated part (due to much larger integration range), we can make the following statement: Given the constraints on $\Omega_m$ from Pantheon+ SNIa of \cite{Brout2022} ($0.334\pm0.018$), we find that we can exclude ratios of $\mathfrak{D}/\epsilon < 4.8\cdot 10^4$ at 95\% CL. This would translate for the current bounds from atomic clocks into $\mathfrak{D} < (2.8 \cdot 10^{-4})/h$ at 95\% CL (using the value from BAO of \cite{Alam2021} with $\Omega_m = 0.299 \pm 0.016$ would give instead $\mathfrak{D} < (3.0 \cdot 10^{-4})/h$ at 95\% CL).\footnote{To obtain these constraints we used the corresponding Gaussian probability densities for $\Omega_m$ and $\epsilon$ from the respective measurements and propagated them according to the laws of transforming and multiplying independent random variables.}

This bound is expected to apply maximally for physically motivated and consistent models emerging from unification theories (as GUTs or string theories), in which the potential contribution is usually subdominant and in which the presence of couplings with radiation freeze the field during radiation domination and typically just lead to a logarithmically slow variation in matter domination, leaving even less time for the field to decelerate.
As we will see below, the presence of a potential that is not fine-tuned does not weaken these bounds.

\newpage
\subsection{The issue of potentials}
We have considered only the homogeneous part of the field evolution of \cref{eq:kg_sol} so far, and one could wonder if the inhomogeneous part involving the potential slope $\d V/\d \phi$ could impart sufficient deceleration to allow the field to evade this issue. Indeed, with complete freedom over initial conditions and shape of the potential, this is always trivially possible. One can simply construct such a potential where $\dot{\phi} \to 0$ towards $z=0$. As we will argue below, this requires a fine-tuned and 'unnatural' combination of initial conditions and potential shape.

Let us consider the energy conservation equation (or equivalently \cref{eq:kg_eom}), which can be written as
\begin{equation}\label{eq:energy_nonconservation}
    \frac{\d V}{\d \ln a} + \frac{1}{2} \frac{\d \dot{\phi}^2}{\d \ln a} = - 3\dot{\phi}^2 < 0~.
\end{equation}
This equation corresponds to motion in a potential with a dissipative force,\footnote{This might become more obvious by replacing $\phi \to \sqrt{m} x$, which then immediately gives the normal kinetic term of a free particle ($\rho = V+\frac{1}{2}m\dot{x}^2$). The dissipative force in this case would simply be $F_\mathrm{diss}=-3H\dot{x}$ (whereas the normal force generated by the potential is still $F_\mathrm{pot} = -\partial_x V$).} and as such the field must eventually obey $\d V/\d \ln a < 0$ (as the squared field speed trivially cannot decrease below $0$). Indeed, at any location within the potential a field would naturally start rolling downhill ($\dot{\phi}=0 \Rightarrow \ddot{\phi}=-\d V/\d \phi$ and thus $\delta [\dot{V}] = - (\d V/\d \phi)^2 \delta t$). As such, we have a strong expectation that $\dot{V} < 0$ except in a transitory phase where the field has been accelerated before and suddenly encounters an uphill slope and has not sufficiently decelerated. We stress here that it is not impossible to generate $\dot{V} > 0$, it is simply not a 'natural' state for a field with arbitrary initial conditions in an arbitrary potential. 

However, the problem is slightly worse than just requiring $\dot{V}>0$ during part of the evolution. This is because \cref{eq:kg_sol} weighs the contributions of a given $\d V/\d \phi$ with the current Hubble expansion rate as well as a factor of $a^3$, making the weight a function that is quite spiked towards $a \to 1$. Indeed, during matter domination we have $a^3/H \propto a^{4.5}$. This implies that the deceleration has to happen mostly towards the end of the evolution, and as such be sudden. Since we are essentially requiring the field to come to an abrupt halt, it does not move over great distances of the potential, thus requiring essentially the potential slope at a singular point (of the current-day field position) to balance exactly to cancel out any remaining velocity. This is the fine-tuning issue for non-oscillatory solutions. In \cref{ssec:tanh_example} we show an example with a constant slope $\mathrm{d}V/\mathrm{d}\phi$ to decelerate the field, and note that the final field slope has to be extremely fine tuned. 


Another way to quickly decelerate the field might in principle use strong oscillations. However, it can quickly be shown that in a potential of type $V(\phi) = A \cdot \phi^{2n}$ the \emph{envelope} of the oscillations only decays as $a^{-3/(n+1)}$\, (see for example \cite{Poulin2018}), and the envelope of the oscillation speed decays as $a^{-3n/(n+1)}$. We reproduce for convenience the argument of \cite{Turner1983} in \cref{app:oscillatory_dynamics}, and extend it to the field speed. In any case, this behavior $a^{-3n/(n+1)}$ is \emph{slower} than $a^{-3}$ for all $n \in \mathbb{N}$, implying that the problem for oscillating potentials is equal or even slightly worse, since they are slightly more inefficient at dissipating energy and they cannot use a slope $\d V/\d \phi$ to decelerate, since it will be traversed in both directions during any oscillation. For this reason even oscillatory potentials cannot create a large $\mathfrak{D}$ for a given small bound on $\epsilon$ using the same reasoning as in \cref{eq:analytical}. However, here too there is an extremely fine-tuned way to avoid a straight no-go theorem. If the oscillation happens to reach its maximum just at $z=0$, then the field speed there will be zero, meaning that any bound in $\epsilon$ is trivially satisfied. Given that the field speed needs to be about $1-2$ orders of magnitude smaller today than the expected speed from the decay of the field speed envelope, this requires even more fine-tuning in the field position. Indeed, one can quickly show that for the aforementioned potential a field offset of $\phi = \phi_\mathrm{max} (1-\varepsilon)$ gives a relative velocity offset of $\sqrt{2 n \varepsilon}$. Restricting the velocity to be smaller than the velocity envelope by two orders of magnitude (to reach $\mathfrak{D}/\epsilon \sim 10^7$, which is two orders of magnitude larger than the bound of \cref{eq:analytical}) then would bound $\varepsilon \lesssim 10^{-4}/(2n)$, which especially for large $n$ is quite a fine-tuned position for the final field. We discuss such fine-tuning of oscillatory potentials in an example case in \cref{ssec:ede}

In summary, if the potential is oscillatory an extreme degree of fine-tuning is required for a given value of the fine-structure constant variation $\mathfrak{D}=\Delta \alpha/\alpha$ to be compatible with the strict observations of $\epsilon \ll 1$, namely that the field happens to exactly end up at the turnaround of the oscillation today. If the potential is not oscillatory, one needs to construct a potential that happens to have exactly the correct slope towards $z \to 0$ in order to decelerate the field by just the right amount at the last moment. In either case, the potential and/or the field initial conditions need to be fine-tuned in order to avoid the tight constraints. We show such fine-tuning explicitly for a few examples in \cref{sec:results}.  
\begin{statement}\label{stat:main}
    We have demonstrated that one has to either give up large fine-structure constants at recombination or face an extreme degree of fine tuning of the potential or the initial conditions for any model where the fine-structure constant variation is generated from the variation of a single scalar field.
\end{statement}

\newpage

\section{Example studies}\label{sec:results}
The aim of this section is to illustrate and strengthen the argument made in \cref{stat:main}. For this, we investigate two example cases. One displays a highly oscillatory behavior (due to an axion-like potential) and is investigated in \cref{ssec:ede}, whereas the other shows a fine-tuned potential that aims to decelerate the field rapidly towards the end of its movement and is investigated in \cref{ssec:tanh_example}.

\subsection{Axion-like-particles coupled to electromagnetism}\label{ssec:ede}

Let us now verify the validity of the discussion above by considering the canonical case of axion like particles (ALP). Originally invoked to solve the strong CP problem in quantum chromodynamics \cite{Peccei2008}, the existence of ALP is commonly proposed in cosmology as a dark matter candidate \cite{Adams2022}, a source of early dark energy \cite{Poulin2018} or at the origin of cosmic birefringence in the CMB \cite{Minami2020,Diego-Palazuelos2022,Diego-Palazuelos2023}.

The potential $V_n(\phi)$ of an ALP is typically modeled by the function
\cite{axion-dark-energy,axion-cosmology}:
\begin{equation}\label{eq:potential_alp}
    V_n(\phi)= (m_a f_a)^2\left[ 1 - \cos\left(\frac{\phi}{f_a}\right) \right]^n,
\end{equation}
which allows for a domination of the field energy density at early times followed by a brutal damping in successive era. As such, ALP are well motivated candidates to implement an early variation of the fine-structure constant, assuming their full coupling to electromagnetism as proposed in \cite{Flambaum2009}.

The eventuality of coupling between EDE and electromagnetism was already investigated in a different context by \cite{Calabrese2011}, and in some sense we are generalizing here this work for axionic fields and with new datasets.

\addtolength{\textheight}{2\baselineskip}
One can further note that such a phenomenology could be well motivated within the framework of string theory~\citep{Alexander2019}, where both EDE and the early variation of the fine-structure constant could be caused by the existence of a dilaton field coupled to an axion. We will however not discuss this case here, focusing only on single scalar fields models for now. As such, we postpone such an investigation for future work.

The 'canoncial' ALP early dark energy model that has been proven to have a significant impact on the Hubble tension \cite{Poulin2018,Poulin2019,Poulin2023,H0olympics} has four fundamental parameters describing the model. These are $\{m_a, f_a, n, \phi_i\}$, where the first three parameters describe the potential of \cref{eq:potential_alp} and the fourth parameter is the initial condition of the field displacement (the initial field velocity is typically irrelevant due to the strong Hubble drag at early times). These parameters are typically further replaced by parameters carrying more physical meaning, such as $f_\mathrm{ede}$ (the largest fraction of energy density that the ALP EDE field reaches, related to $f_a$) and $z_c$ (the critical redshift at which the field begins oscillating due to the weakening Hubble drag, related to $m_a$). 

Furthermore, for the sake of simplicity, the initial field displacement is replaced by the initial position along the cosine curve of the potential, $\Theta_i = \phi_i / f_a$\,. In addition to this parameter space, we introduce the linear coupling~$\zeta$ of the fine-structure constant to the field displacement as in \cref{eq:delta_alpha}, self-consistently propagating the effect of a varied fine-structure constant throughout recombination processes as described for example in \cite{Hart:2017ndk}. As such, the parameter base can be written as $\{z_c, f_\mathrm{ede}, n, \Theta_i, \zeta\}$ in this notation. 

To investigate the behavior of an ALP EDE coupled to electromagnetism, we used a modified version of the {\sc class} software \cite{Lesgourgues:2011CLASS} coupled to {\sc Montepython}\footnote{\url{https://github.com/brinckmann/montepython_public}} \cite{Brinckmann:2018}, along the line of previous works using a similar setup \cite{Vacher2022,Vacher2023,Schoneberg2023} (following also the idea behind the implementation of \url{https://github.com/PoulinV/AxiCLASS} \cite{Smith:2019ihp,Poulin2018}, but adapting it to the new coding standard of \texttt{CLASS~v3.2.0}, which also involves variations of fundamental constants\footnote{Our implementation of an \emph{instantaneous} transition of the fine-structure constant is publicly available on the software's repository: \url{https://github.com/lesgourg/class_public}.}). For the plots we use \texttt{liquidcosmo}\footnote{Available at \url{https://github.com/schoeneberg/liquidcosmo} based on getdist \url{https://github.com/cmbant/getdist}.}. For simplicity, we first focus on the $n=3$ case, which was favored by observational data in order to address the $H_0$ tension \cite{Poulin2019}.

As further discussed in \cref{app:alpha-models}, the model acts as desired for a model implementing an early variation of the fine-structure constant, inducing almost no variation of the fine-structure constant at low redshift, until it suddenly reaches a plateau associated with a different value of $\alpha\neq \alpha_0$ after an oscillatory transition close to the CMB epoch. The time at which this transition occurs is directly given by the critical redshift ($z_c$). The parameters $f_\mathrm{ede}$ and $\zeta$ impact the amplitude of the oscillations and the magnitude of the plateau in  $\alpha(z)$ while $\Theta_i$ impacts the number of oscillations during the transition phase and the final speed of the field.

In order to confront this model with data, we use likelihoods associated to various datasets\footnote{We do not discuss the constraints coming from Big Bang Nucleosynthesis here (which would further reinforce the point we want to make), as these are expected to be strongly model dependent. On this topic, see e.g. \cite{ClaraMartins2020BBN}.} also used in \cite{Vacher2023}, namely
\begin{enumerate}
    \item \label{it:cosmo} {\textbf{Cosmological:}} we use CMB angular power spectra and lensing reconstruction data from the Planck satellite \cite{Planck_lkl,Planck2018VIII}, baryonic acoustic oscillation data from BOSS DR12 \citep{DR12}, the Pantheon SNIa sample \cite{Scolnic2022}, and cosmic chronometers from \cite{Moresco2022}. We also use a prior on the $H_0$ value from \cite{Riess2021}, implemented as discussed in \cite{Camarena2021} as a prior on the supernovae absolute magnitude.
    \item \label{it:fine-structure} {\textbf{Fine-structure:}} we use the spectroscopic measurement points of quasars (QSO) from various datasets \cite{Martins2017review,alphaWebb,alphaSubaru,alphaespresso} as well as a prior from the Oklo natural nuclear reactor \cite{Oklo}, and the updated bound on the time variation of $\alpha$ of \cref{eq:atomicclocks} coming from measurements of atomic clocks \cite{Filzinger2023}, as already used in \cite{Schoneberg2023}.
\end{enumerate}

\begin{figure}
    \centering    \includegraphics[width=0.45\textwidth]{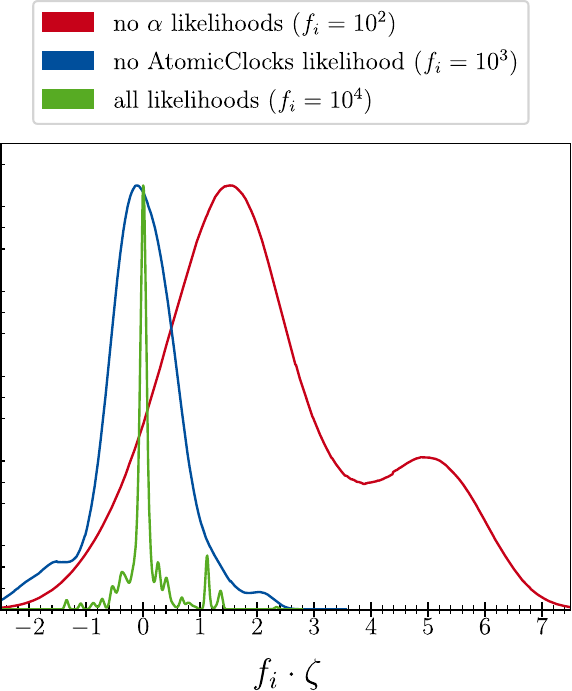}    
    \caption{Contour plots of the parameter space of the ALP scenario for three different data set: no $\alpha$ data (red), no atomic clock data (blue) and all likelihoods (green). $f_i$ is a rescaling factor allowing to easily compare the curves.}
    \label{fig:axion_data_zeta}
\end{figure}

\begin{figure*}[!t]
    \centering    \includegraphics[width=0.6\textwidth]{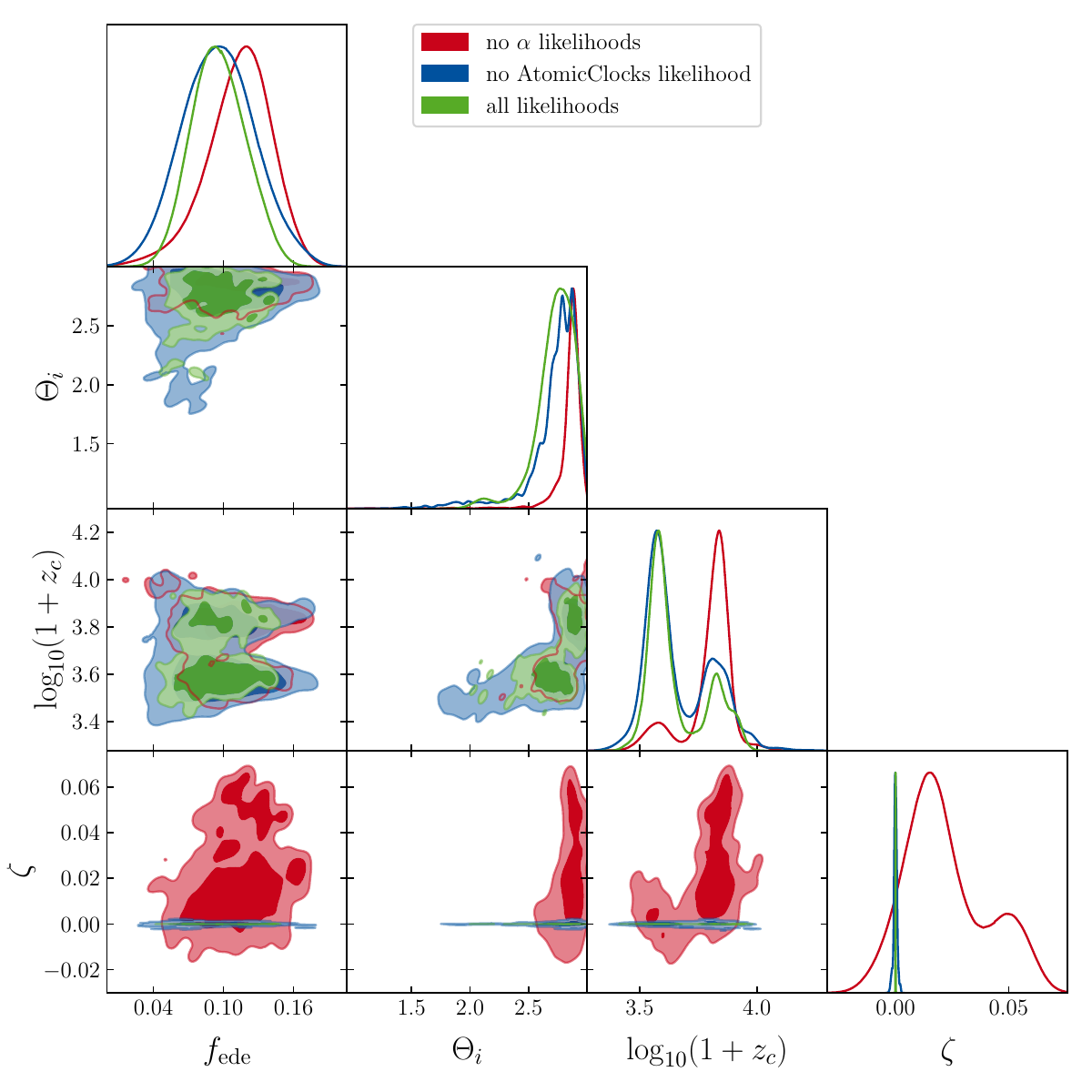}    
    \caption{Contour plots of the parameter space of the ALP scenario for three different data set: no $\alpha$ data (red), no atomic clock data (blue) and all likelihoods (green).}
    \label{fig:axion_data}
\end{figure*}

The results of this investigation can be found in \cref{fig:axion_data}, where we show the constraints on the underlying model parameters from only cosmological observations (\cref{it:cosmo}) or also including fine-structure constant observations (\cref{it:fine-structure}). A zoom-in on the constraints on $\zeta$ can be found in \cref{fig:axion_data_zeta}. The immediate conclusion is that without experiments probing the fine-structure constant, a large range of $\zeta$ values is allowed (even slightly favoring a positive value) For example, a point on the edge of the $2\sigma$ exclusion limit with $\zeta\simeq 0.06$ can be associated with a value of $\Delta\alpha(z_{\rm cmb})/\alpha_0\sim 10^{-3}$.\footnote{These suggestive values of $\Delta\alpha(z_{\rm cmb})/\alpha_0$ naturally depend on the other parameters of the model (such as $\Theta_i$ or $z_c$). Here we use the bestfit values of the corresponding run -- rescaling only $\zeta$ -- to give a qualitative estimate.}

Instead, once likelihoods probing the fine-structure constant are added, no such points are available anymore, and the $\zeta$ is restricted to lead to cosmologically irrelevant $\alpha(z)$ at the recombination times, in accordance with our almost-no-go theorem of \cref{stat:main}. Put otherwise, while it would be legitimate to think that local constraints on $\alpha(z)$ at $z\sim 0$ have no bearing on the possible behavior of the field during recombination era at $z\sim 1100$, quite the opposite is true for a well-defined single-field model. With the full likelihood set, the maximum values allowed for the electromagnetic coupling are of the order of $\zeta\sim 10^{-4}$, which can be associated to a value of $\Delta\alpha/\alpha_0(z_{\rm cmb})\sim 10^{-6}$.

Overall, we report constraints on $\zeta$ and the associated suggestive values of $\Delta\alpha/\alpha_0(z_{\rm cmb})$ in \cref{tab:results}.\footnote{We note that the ALP EDE model is always what drives the higher value of $H_0$ in this combined model (and there is no correlation between $\zeta$ and $f_\mathrm{ede}$) since the EDE is so much more efficient at resolving the Hubble tension in a way that is statistically preferred from the CMB data compared to the $\alpha(z)$ introduced by the coupling ($\zeta$).}

\begin{table}[t]  
\caption{\label{tab:results}Mean and standard deviation of $\zeta$ for the various data combinations in the ALP EDE case, corresponding to the posteriors shown in \cref{fig:axion_data_zeta} and corresponding variation of the fine-structure constant at the emission of CMB associated to the mean.}
  \centering
  \resizebox{0.5\textwidth}{!}{
  \begin{tabular}{c|c|c|c}
    \toprule
           & No $\alpha$ data & No Atomic clocks & All likelihoods \\ \midrule
    $\zeta$ & $(2.2 \pm 1.9) \cdot 10^{-2}$  & $(-0.06\pm0.77)\cdot 10^{-3}$ & $(0.06\pm0.45)\cdot 10^{-4}$  \\
    \midrule
    $\frac{\Delta \alpha}{\alpha_0}\left(z_{\rm cmb}\right)$ & 
    $2.5\cdot 10^{-4}$  & $-6.8\cdot 10^{-6}$  & $6.8\cdot 10^{-8}$ \\
    \bottomrule
  \end{tabular}
  }
\end{table}


Importantly, it is possible to circumvent the relatively narrow bounds on $\zeta$ reported in \cref{tab:results} entirely if one is allowed to fine-tune their initial conditions ($\Theta_i$) drastically, such that the final field movement will be exactly in a turnover of an oscillation. Since these fine-tuned initial conditions take up a tiny negligible prior volume, they are not visible in the posterior contours of \cref{fig:axion_data}. This behavior of fine-tuned initial conditions or potential parameters occupying a vanishing amount of prior space is expected to be generic and not just related to this particular model.


This can lead us to conclude another interesting corollary statement: 
\begin{statement}\label{stat:bayes_vol}
    Even if fine-tuned initial conditions and/or potential shape can reconcile large variations of the fine-structure constant at recombination with the local bounds, these typically occupy a negligible fraction of the prior volume and might thus still be ruled out in a Bayesian analysis.
\end{statement}
However, in our ALP EDE example case we can force the field into such fine-tuning by minimizing the likelihood for a fixed $\zeta = 10^{-3}$ (more than an order of magnitude larger than the allowed range). In this case, we can reach a similarly good maximum likelihood as in the full run, differing only by $\Delta \chi^2 \simeq 23.1$ (largely driven by the QSO and Oklo likelihoods), whereas the original bestfit of the full run would result in a $\Delta \chi^2 = 129\,000$ if simply rescaled to have $\zeta=10^{-3}$. The reason is that with this minimization strategy, the fine-tuning of the initial conditions can be performed such that the final field configuration is almost perfectly at the turnover of a single oscillation. 

\begin{figure}
    \centering    \includegraphics[width=0.45\textwidth]{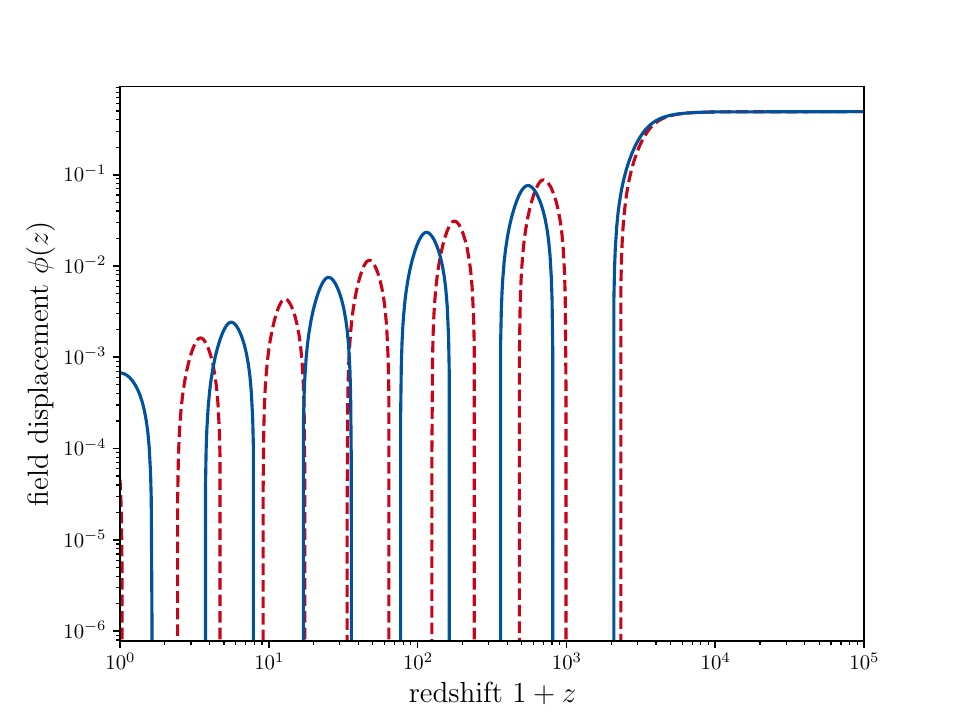}    \includegraphics[width=0.45\textwidth]{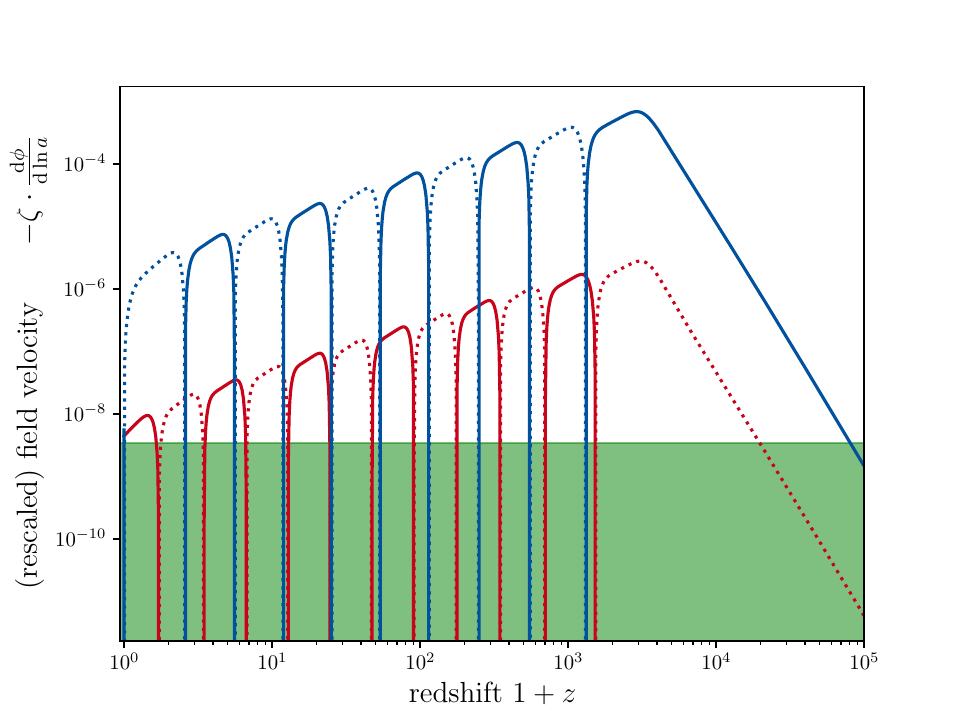}
    \caption{Axion EDE field (top) and field velocity (bottom) as a function of redshift for the bestfit of the full model (red) and for the bestfit of the model with $\zeta = 10^{-3}$ fixed (blue). While the red model reaches a small rescaled field velocity due to a small value of $\zeta$, the blue model reaches it only through the fine tuning of the initial conditions allowing for the exact cancellation of the field speed today in the turn-over of an oscillation.}
    \label{fig:field_and_velocity}
\end{figure}

\newpage
Such a behavior is displayed in \cref{fig:field_and_velocity}, comparing the minimized bestfit of the full run and the minimized bestfit of the run with $\zeta=10^{-3}$ fixed. While the former reaches the small final field velocity through appropriately small $\zeta$ to lead to cosmologically irrelevant $\alpha(z)$ at recombination times, the latter reaches it by fine-tuning the $\Theta_i$ such that the final field-velocity is almost zero by virtue of the oscillation ending at exactly $z=0$ (see also the field displacement being at a peak in the same \cref{fig:field_and_velocity}). 

Another interesting corollary to the main \cref{stat:main} can be derived from the fact that we cannot find a region of $\Delta \chi^2 \simeq 0$: since this 'trick' of fine-tuning the deceleration of the field at late times cannot lead to vanishing $\Delta \alpha(z)$ for a wide range of redshifts in an oscillatory potential, the QSO and Oklo likelihoods at $z>0$ necessarily impart a reasonably large likelihood penalty. 
\begin{statement}\label{stat:other_data}
    As such, even when there is an extreme degree of fine-tuning in the design of the potential, other data on the fine-structure constant in the late Universe can put even further constraints against a relevant fine-structure at recombination.
\end{statement}
\vspace{2\baselineskip}

\subsection{Toy model}\label{ssec:tanh_example}

\begin{figure}
    \centering
    \includegraphics[width=0.45\textwidth]{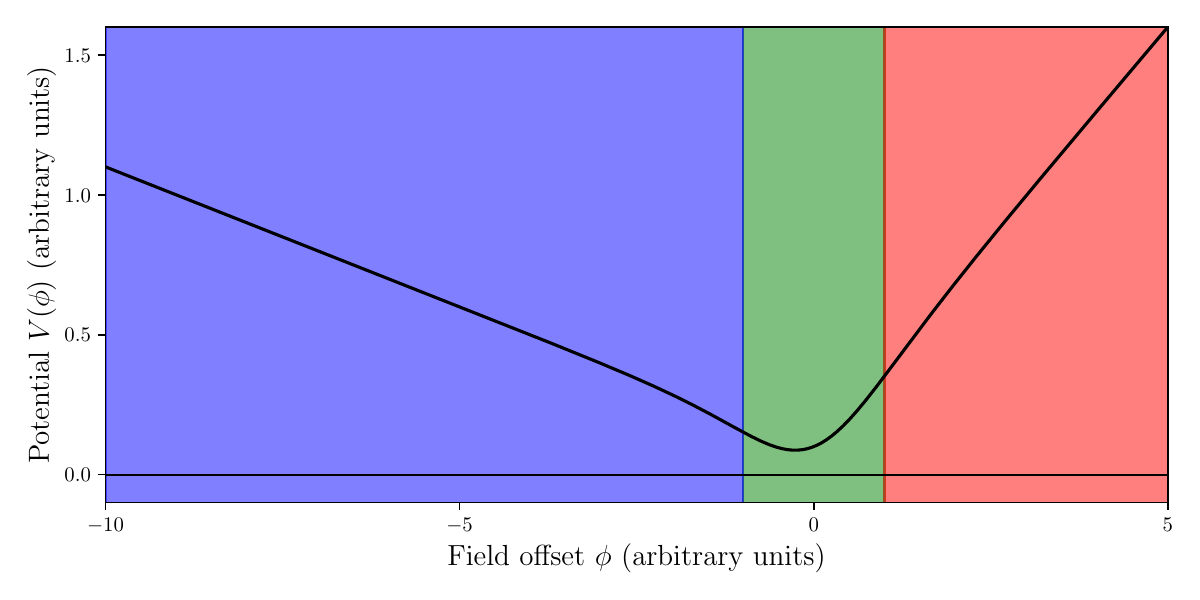}
    \caption{Example of a potential of the form of \cref{eq:zth_potential} with parameters $a=0.3$, $s=-0.1$, $\Sigma=1$, $\kappa=0$, and $V_0 = 10^{-11}$. The blue range marks the accelerating part of the potential dominated by the downhill slope $s<0$ for $\phi \ll -\Sigma$, while the red marks the sudden deceleration regime dominated by the uphill slope $a>0$ for $\phi \gg \Sigma$. The intermediate region is marked in green.}
    \label{fig:zth_potential}
\end{figure}

In this section we attempt to build a simple toy model of a case in which the field does not oscillate around a minimum in the potential, but instead displays a continuous motion. As such, according to \cref{stat:main} it can only have a significant impact at the time of recombination and avoid the current fine-structure constraints by virtue of fine-tuning of initial conditions/parameter values. For this, we introduce a potential based on the hyperbolic tangent function as

\begin{equation}\label{eq:zth_potential}
    V(\phi) = 10^{-10} \cdot [ f a \phi + (1-f)(s\phi + \kappa)] + V_0
\end{equation}
with
\begin{equation}    
f=\frac{1}{2}\left(\tanh\left(\frac{\phi}{\Sigma}\right)+1\right)~,
\end{equation}
with the free parameters $\{V_0, \kappa, a, s, \Sigma\}$\footnote{The offset parameter $\kappa$ was introduced here for the sake of generality but does not display any interesting behavior.}. The factor $10^{-10}$ is used purely for numerical reasons related to the implementation in \texttt{class}. We show one example of such a potential in \cref{fig:zth_potential}. 

The idea of this potential is to first accelerate the field down a slope, which is accomplished by the term $s \phi$ with a negative slope $s<0$, relevant as long as $\phi \ll -\Sigma$. As soon as the field crosses the inflection point ($\phi \gg \Sigma$), the second part of the potential becomes important and decelerates the field through another linear slope $a \phi$, now with positive slope $a>0$. The value of $\Sigma$ quantifies the width of the transitional region. We additionally allow for an offset $\kappa$, which could in principle further accelerate/decelerate the field only within the transitional region.

As further discussed in \cref{app:alpha-models}, this toy model provides a great example of early fine-structure constant variation, with a single brutal transition of the value of $\alpha$ around $z_{\rm cmb}$, very close to a Heaviside function. The two slopes $s$ and $a$ together drive the time and the amplitude of the transition. As expected, $\zeta$ simply rescales the whole evolution of the fine-structure constant.

For our toy experiment, we fix the cosmological parameters to the best fitting parameters of Planck \cite{Planck2018} and investigate only the extension of the parameter space, using flat priors on $\log_{10}(-s)\in[1,15]$, $\log_{10}(\kappa)\in[-5,5]$, $\log_{10}(a) \in [-10,5]$, and $\zeta\in[-10,10]$ and we fix the parameters $\Sigma=10^{-5}$ and $V_0=10^{-8}$. The initial field value is also fixed to $\phi(z\to\infty)=-10^{-2}$. Since the parameter space is highly degenerate and difficult to explore with traditional MCMC methods, we make use of the {\sc polychord} sampler\footnote{\url{https://cobaya.readthedocs.io/en/latest/sampler_polychord.html}} in this case.

\begin{figure}
    \centering    
    \includegraphics[width=0.45\textwidth]{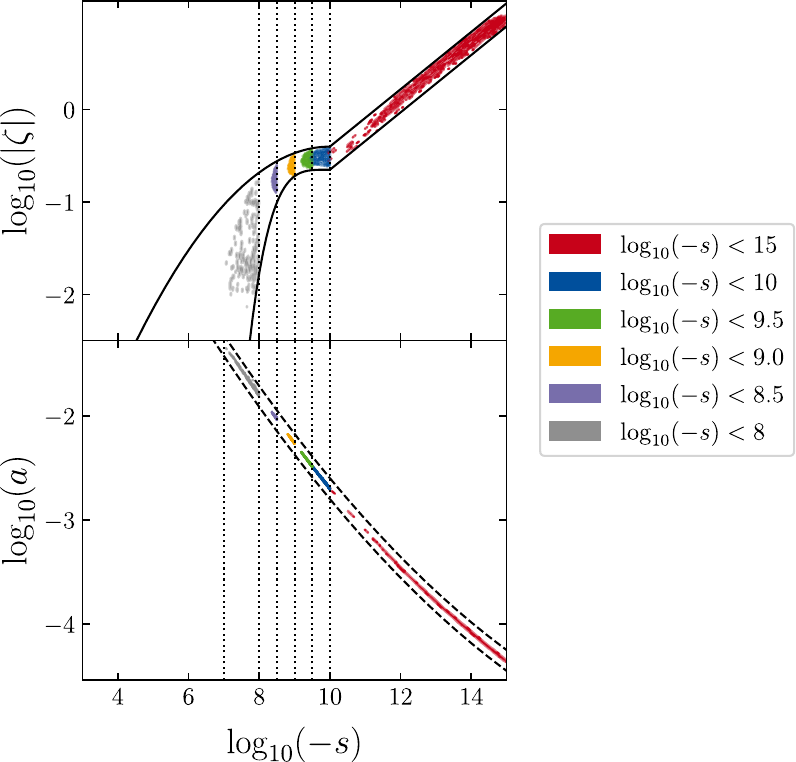}    
    \caption{Constraints on the parameter space of the toy model with the hyperbolic tangent potential. Countours are derived for different bounds on $\log_{10}(-s)$.}
    \label{fig:many_zth}
\end{figure}

\begin{figure}
    \centering
    \includegraphics[width=0.45\textwidth]{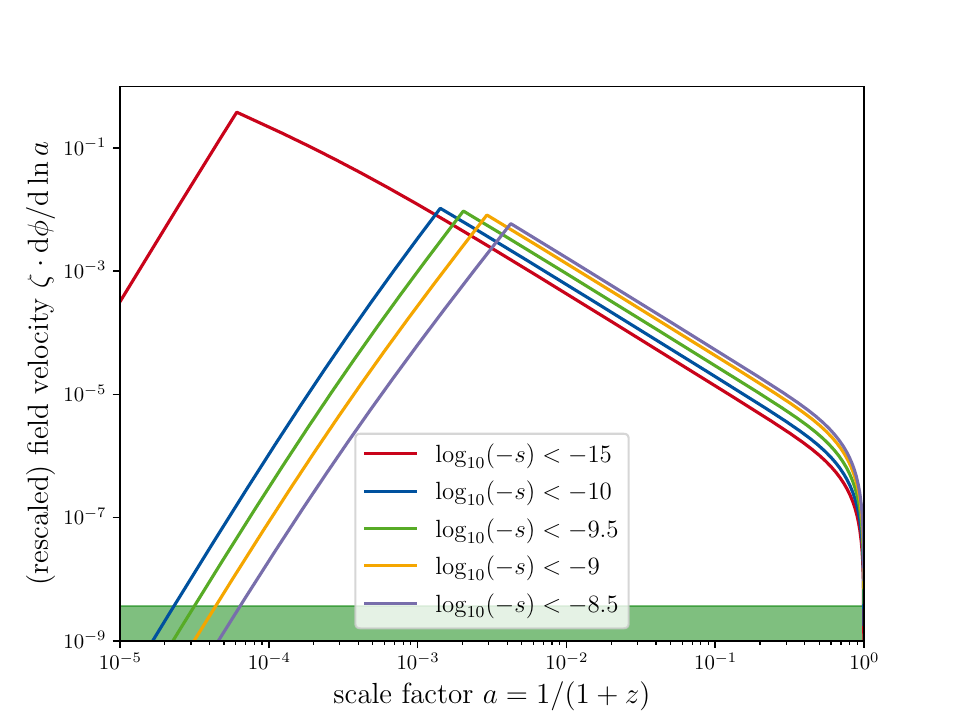}
    \caption{Rescaled field speeds for the best fitting models for the various cases. Note that cases with lower $\log_{10}(-s)$ need a smaller $\zeta$ to reach the same low rescaled field velocity (the peak of the non-rescaled velocity is always at around $\mathcal{O}(10^{-1})$). All of the models show a very drastic deceleration towards $z\to 0$ or equivalently $a \to 1$.}
    \label{fig:zth_velocity}
\end{figure}

We only put two requirements on the field, namely that it should generate a cosmologically relevant variation of the fine-structure constant at recombination, $\Delta \alpha(z_\mathrm{cmb}) = 10^{-2} \pm 10^{-3}$ (thus avoiding the potential issue discussed in \cref{stat:bayes_vol}) and that it should avoid constraints by the fine-structure observations detailed in \cref{it:fine-structure}. These two limitations alone are sufficient to force the model parameters into a tight fine-tuned degeneracy, which can be observed in \cref{fig:many_zth} (lower panel). The degeneracy manifests as a one-to-one relation between the slope for acceleration ($-s$) and the slope needed for deceleration ($a$). This slope for deceleration has to balance exactly in such a way as to make the field decelerate entirely until today. Examples of the field trajectories are shown in \cref{fig:zth_velocity} where this becomes very apparent as a sudden and rapid stop/drop in velocity towards $z \to 0$ or equivalently $a = 1/(1+z) \to 1$. 

It should also be noted that by construction the field speed typically reaches $\mathcal{O}(10^{-1})$ for this type of potential, thus requiring smaller values of $\zeta$ when the field decelerates later, which is typically the case for smaller slopes. This imparts an additional (though not as strong) degeneracy between $\log_{10}(|\zeta|)$ and $\log_{10}(-s)$ that can be seen in \cref{fig:many_zth}.
As such, even in this case we can conclude that one has to fine-tune the potential such that the field suddenly and rapidly decelerates exactly at $z \to 0$ if one requires cosmologically relevant $\Delta \alpha(z)$ at the time of recombination as per \cref{stat:main}.
\newpage

\section{Discussion and Conclusions}
\label{sec:conclusion}

In this work we have theoretically derived an almost-no-go theorem (\cref{stat:main}) about the impossibility to have a large fine-structure constant at recombination while remaining compatible with late time observations in a single field model without resorting to extreme fine tuning of the potential or its initial conditions. 

We have given two examples in \cref{sec:results} that show the impact of this almost-no-go theorem. Furthermore, we have derived two collary statements to this main statement, which further restrict the fine-tuned models that avoid \cref{stat:main}. \Cref{stat:bayes_vol} states that the prior volume of such fine-tuned models in a Bayesian analysis is typically negligible, while \cref{stat:other_data} states that other late-Universe data probing the fine-structure data can still be an issue in this case (especially for oscillating potentials). As such, it appears that the often-investigated possibility of a fine-structure constant variation around the time of recombination that has been proposed in the past to ease the Hubble tension \cite{Chluba2023} does not seem feasible with current late-time and laboratory observations of the fine-structure constant, at least for single-field models.

\enlargethispage{1\baselineskip}
Indeed, using fairly simple arguments one can derive an upper bound on the variation of the fine-structure constant at recombination of $\Delta \alpha(z_\mathrm{cmb})/\alpha_0~<~3~\cdot~10^{-4}/h$ (95\%~CL), which can be further generalized to $\Delta \alpha(z_\mathrm{cmb})/\alpha_0 < 5 \cdot 10^{-4}$ using a weak prior on the Hubble parameter. 

We note that there are a few caveats to this discussion. First and foremost, the argument can be avoided if screening mechanisms exist that would either allow the fine-structure constant $\alpha$ to take on a different value depending on the scale at which it is measured (such as the chameleon mechanism, see for example \cite{Olive2008}) or that would act differently on the values of the fine-structure constant in the late and the early Universe.\footnote{Screening mechanisms would typically imply a change in the effective value of $\alpha$ with the local density and hence discard laboratory measurement such as atomic-clocks, the density on the Earth's surface being much greater than the one of the primordial plasma at recombination. This would however leave the measurements of QSO available, which derive from absorption lines in low density clouds. However, using the bounds of $(\mathrm{d}\alpha/\mathrm{d}t)/\alpha_0 |_{z\simeq1.5}\sim 10^{-16}h /\mathrm{yr}$ from QSO measurements and using the same reasoning as above only lead to the mild upper bound of $\Delta \alpha(z_\mathrm{cmb})/\alpha_0 \lesssim 0.1$.} Second, we perform a linearization in the field displacement, the validity of which is further discussed in \cref{app:linearization}. Third, our model does not consider possible couplings of the field to the other sectors of the Universe (such as baryonic matter), as is the case in many theories involving a varying $\alpha(z)$. The existence of such couplings, as the Bekenstein-type couplings $\zeta_i\rho_i$ \cite{Olive2002}, should impact slightly cosmic evolution of the field but we do not expect it to drastically change our conclusions (the field deceleration still needs to be fine-tuned in terms of the additional coupling parameters).

We also want to point out that the same argument can in principle be made for the variation of other fundamental constants, such as the electron-to-proton mass ratio, which has been shown in the past to be even more successful in easing the Hubble tension \cite{H0olympics,Lee2023,Sekiguchi2021}. Ultimately, such an investigation would have to consider the combined variation of the fundamental constants that is expected in most physically motivated models or define a well-defined theory of varying only the electron mass.

Let us also note here that any varying constant model must induce a violation of the Einstein equivalence principle at some level (see for example \cite{Damour2002,Uzan2011}), such that it should also be sharply constrained locally by stringent tests of the universality of free fall performed by experiments such as MICROSCOPE \cite{microscope}. We did not consider such a constraint here as its relation to the underlying parameters is strongly model-dependent and the accurate bound provided by atomic clocks is in any case stringent enough for our statements.

The improvement brought by future datasets is expected to put further stress on coupled scalar field models. The stringent bounds on the stability of fundamental constants from laboratory data will be sharpened by the upcoming nuclear clocks experiments \cite{Thorium} while early Universe constraints will be updated by future generation of CMB missions such as Simons Observatory \cite{SimonsObservatory} and the combination of CMB-S4 \cite{CMBS4} and LiteBIRD \cite{PtepLB}, which will measure the anisotropies of the last scattering surface at a cosmic variance accuracy.\footnote{See \cite{Hart:2021kad,Tohfa2023} as well as \cite{Hart2023} for quantitative forecasts on constraints on varying constants coming from measurements of anisotropies and spectral distortions with future CMB experiments. The combination of these datasets 
will rule out any cosmologically interesting variation of the fundamental constants at recombination, and thus condemn simple and consistent solutions of the Hubble tension based on such variations.
}

While the scope of the almost-no-go theorem has to be clearly defined, we stress that the impressive work that has gone into tightening the local laboratory constraints on variations of fundamental constants have yielded such high precision that now even seemingly unrelated cosmological epochs are constrained by these data.

\section*{Acknowledgments}

The authors would like to thanks C.J.A.P. Martins,  Savvas Nesseris and J.F. Dias for their precious comments on the final version of the draft.

LV acknowledges partial support by the Italian Space Agency LiteBIRD Project (ASI Grants No. 2020-9-HH.0 and 2016-24-H.1-2018), as well as the the RadioForegroundsPlus Project HORIZON-CL4-2023-SPACE-01, GA 101135036. This work was partially done in the framework of the FCT project 2022.04048.PTDC (Phi in the Sky, DOI 10.54499/2022.04048.PTDC).

Computations were made on the Mardec cluster supported by the OCEVU Labex (ANR-11-LABX-0060) and the Excellence Initiative of Aix-Marseille University - A*MIDEX, part of the French “Investissements d’Avenir” program. LV would like to thanks B. Carreres for several helps with the use of Mardec.

NS acknowledges the support of the following Maria de Maetzu fellowship grant: Esta publicaci\'on es parte de la ayuda CEX2019-000918-M, financiada por MCIN/AEI/10.13039/501100011033. NS also acknowledges support by MICINN grant number PID 2022-141125NB-I00.

\bibliography{apssamp}
\appendix
\section{Oscillatory dynamics}\label{app:oscillatory_dynamics}
Note that this is a simplified write-up of \cite{Turner1983}. The interested reader is encouraged to visit the original source.

In order to determine the behavior of the field envelope, the idea is that the field oscillates between the two critical points of the potential. These are the points where simultaneously $\dot{\phi}=0$ and $V(\phi_\mathrm{max})=V_\mathrm{max}$, with $V_\mathrm{max}$ being relatively constant over a single oscillation (as long as the oscillation frequency is much larger than the Hubble frequency).

Then the evolution of $V_\mathrm{max}$ is directly a tracer of the total energy density $\rho$ (since $\rho(\phi_\mathrm{max}) = V_\mathrm{max}$ by definition). The advantage of using $\rho$ instead of $V$ is that $\rho$ will typically not decrease drastically during the oscillation, as the potential energy is turned into kinetic energy. To find the density evolution, we decompose $\dot{\phi}^2 = \rho+P \equiv (\gamma+\gamma_p) \bar{\rho}$ with some constant part $\gamma$ and some oscillatory part $\gamma_p$ and some mean evolution $\bar{\rho}$. Note that the energy density is \emph{not} conserved (see \cref{eq:energy_nonconservation}), though the overall stress-energy is. Indeed, we use the energy conservation equation $\d \rho/\d \ln a = -3(\rho+P) = -3 (\gamma+\gamma_p) \bar{\rho}$ (which is equivalent to \cref{eq:energy_nonconservation}) to find an equation involving $\bar{\rho}$. On the long timescales of interest we neglect $\gamma_p$ (average it out, giving also $\bar{\rho}$ from the average of $\rho$), and we are left with $\bar{\rho} \propto a^{-3 \gamma}$. To explicitly find $\gamma$, one can average $\dot{\phi}^2/\bar{\rho} \equiv \gamma+\gamma_p$ over any given single oscillation (which will be the same in any other oscillation) and neglect for this the slow variation of the field speed due to the Hubble friction. In this limit $\dot{\phi}=\sqrt{2(\rho-V)}$. This gives a period of oscillation of 
\begin{equation}
    T = \int \frac{\d t}{\d \phi} \d t = \int_{-\phi_\mathrm{max}}^{\phi_\mathrm{max}} \frac{1}{\sqrt{2(\rho-V)}}\d \phi
\end{equation}
Plugging everything in, then leaves us with 
\begin{align}
    \gamma &\approx \frac{1}{T} \int_0^T \dot{\phi}^2/\bar{\rho} \d t = \frac{1}{T} \int_{-\phi_\mathrm{max}}^{\phi_\mathrm{max}} \dot{\phi}/\bar{\rho} \d\phi \\ &= \left[\int_{-\phi_\mathrm{max}}^{\phi_\mathrm{max}} \sqrt{2(\rho-V)}/\sqrt{\bar{\rho}}\d \phi\right]\Bigg/\left[\int_{-\phi_\mathrm{max}}^{\phi_\mathrm{max}} \frac{\sqrt{\bar{\rho}}}{\sqrt{2(\rho-V)}}\d \phi\right]
\end{align}
For the final step, we approximate $\rho\approx\bar{\rho}$ during the oscillation and use $V/\bar{\rho} \approx V/V_\mathrm{max} \approx (\phi/\phi_\mathrm{max})^{2n}$ to obtain
\begin{equation}
    \gamma = \frac{2n}{n+1}
\end{equation}
which can then easily be plugged into the equation for 
\begin{equation}
    V_\mathrm{max} \propto \bar{\rho} \propto a^{-3\gamma} \propto a^{-6n/(n+1)}
\end{equation}
and finally using $V_\mathrm{max} \propto \phi_\mathrm{max}^{2n}$ we get the evolution of the envelope as
\begin{equation}
    \phi_\mathrm{max} \propto a^{-3/(n+1)}
\end{equation}
Note that $\d [\phi_\mathrm{max}]/\d t \approx -\frac{-3H}{n+1} \phi_\mathrm{max}$ and is not equal to $[\d \phi/\d t]_\mathrm{max}$ , the envelope of the oscillation speed. For the latter, we simply recall that $\dot{\phi} = \sqrt{2(\rho-V)}$ is maximal when $V \to 0$, and there gives $\dot{\phi} \propto \sqrt{\bar{\rho}} \propto a^{-3n/(n+1)}$ (the same can be found using the Virial theorem).

\section{Linearization}\label{app:linearization}

One may wonder about the legitimacy of the linearization $B_F\simeq 1 + \zeta \Delta \phi$ assumed throughout all this work, especially for the early time evolution of the fields.

To test this linearization, we compare it with the evolution predicted by other well motivated models. The canonical Bekenstein models
\cite{Bekenstein-2,Sandvik2002} predict a variation of the form
\begin{equation}
\frac{\Delta \alpha}{\alpha_0} = e^{-\zeta\Delta \phi}-1
\end{equation}
with $\zeta=-2$. However, such a high value of $\zeta$ is clearly excluded by data for the potential under consideration here, so we generalize it to smaller values of the electromagnetic coupling. 

Another relevant model is the runaway dilaton model \cite{Damour1994}, in which the fine-structure constant evolution is given by \cite{Martins2015}:
\begin{equation}
\frac{\Delta \alpha}{\alpha_0} = 1-\zeta(1- e^{-\Delta\phi})
\end{equation} 
These two models are expected to be well described by our linearization, at least near $\phi=\phi_0$.

\begin{figure}
    \centering    \includegraphics[width=0.5\textwidth]{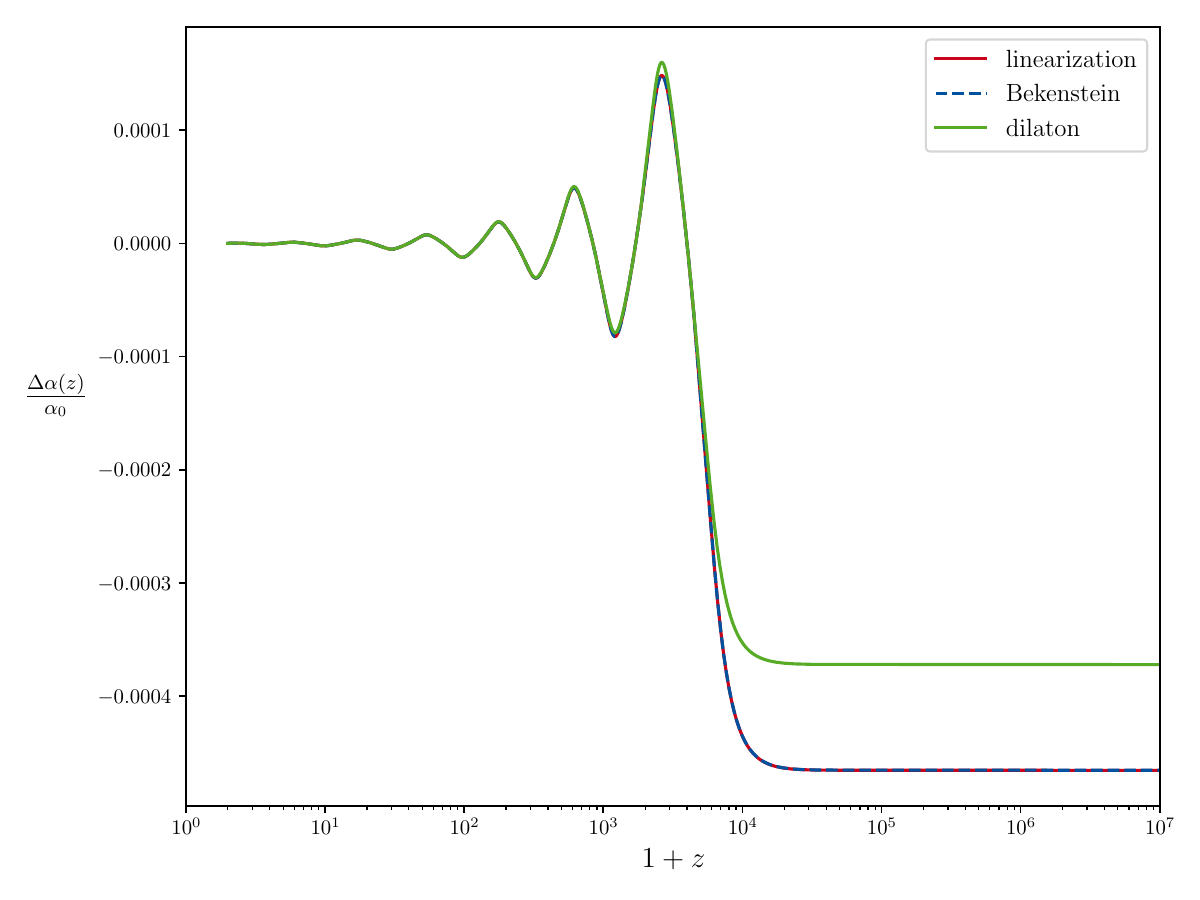}    
    \caption{Comparison of $\Delta\alpha(z)/\alpha_0$ for $\zeta=10^{-3}$ using different expressions: linearization (red), dilaton like coupling (green) and exponential coupling (blue).}\label{fig:linearization}
\end{figure}
The redshift evolutions of $\Delta\alpha/\alpha_0$ in the linearized, the Bekenstein, and the dilaton scenarios are displayed in \cref{fig:linearization} using the axionic potential, $\zeta= 10^{-3}$ and best-fit values for all the other parameters. We can see that the Bekenstein type model is well described by the linearization througout all of the cosmic history with difference never getting greater than one permille. The runaway dilaton prediction however, has the same overall behavior but diverges compared to the linearization (up to $\sim 20\%$) after it reaches a plateau during radiation domination at very high redshift. 

However, this is not a concern for several reasons. First, at the time of recombination ($z \sim 1100$) the typical deviation is still in the permille range. Second, even at earlier times, it can be shown that the leading order correction causes a smaller variation of $\Delta\alpha(z)$ at equal $\zeta$, and thus our fine-tuning argument would be even stronger for these types of couplings. Third, even in other modified coupling scenarios that differ for a constant $\zeta$ by order-unity factors in $\Delta\alpha(z)$, the argument is only slightly weakened but otherwise remains intact due to the large order-of-magnitude between the constraint of \cref{eq:analytical} and the typical $\mathfrak{D}/\epsilon$ required to be cosmologically relevant.

One could even imagine stronger deviations in cases where $B_F(\phi)$ has an even more exotic shape that could not even be linearized. It might thus be possible to counter the argument of this paper by using such a well designed choice of $B_F(\phi)$. However, this choice would have to be motivated from an underlying high energy theory, otherwise it would just amount to displace the extreme level of fine tuning at the level of the intiial conditions/potential shape towards the choice of the $B_F(\phi)$ function itself.

We will not discuss here possible kinetic coupling of the field of the form $B_F(\phi,\partial_\mu\phi)$ as in \cite{Barros2023} and leave such an investigation for future works.

\newpage
\onecolumngrid
\section{Evolution of the fine-structure constant in the models}
\label{app:alpha-models}

\begin{figure*}[!h]
    \centering    \includegraphics[width=0.7\textwidth]{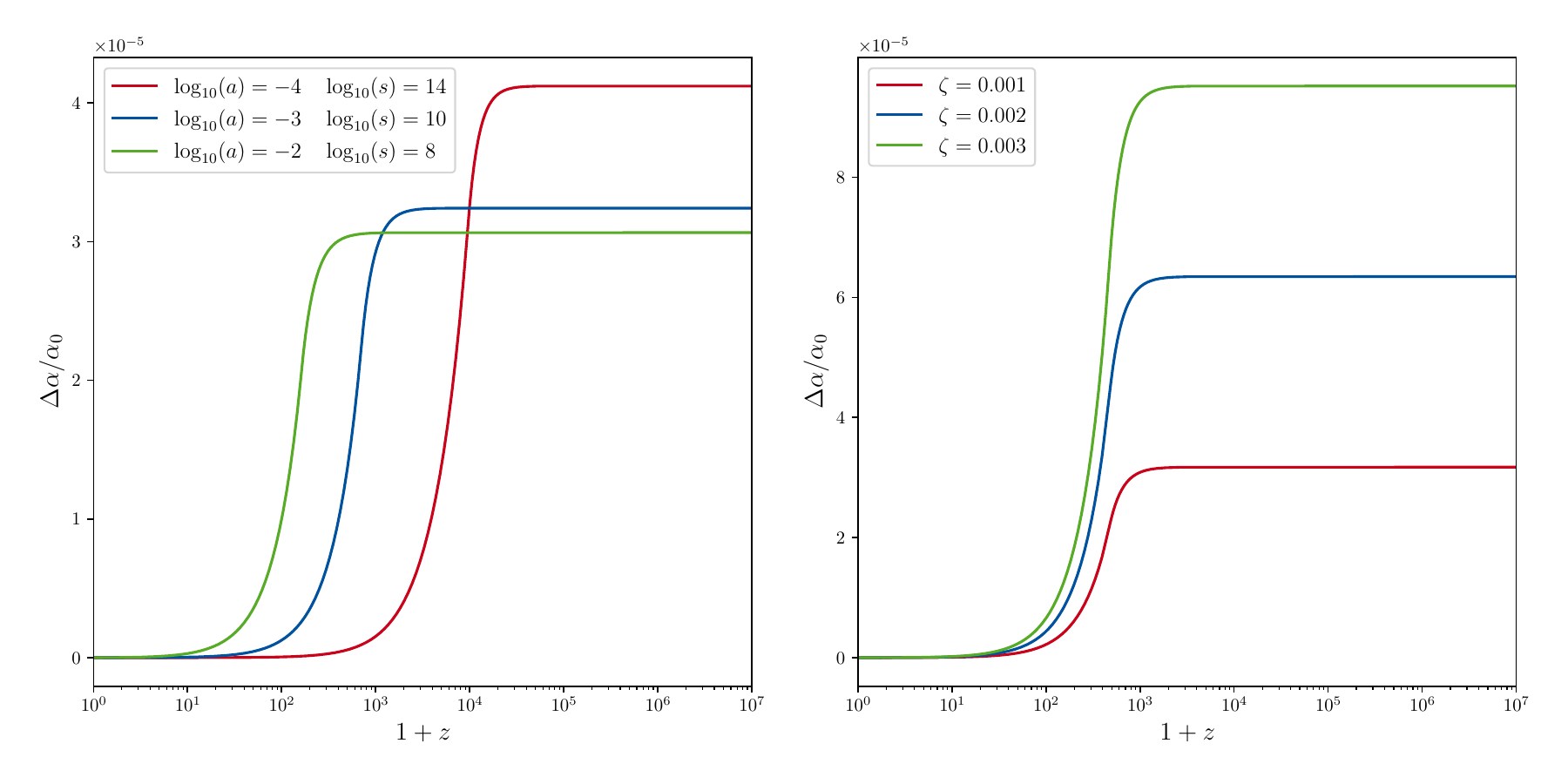}    
    \caption{Impact of the different parameters of the hyperbolic tangent model on the cosmic evolution of $\Delta \alpha(z)/\alpha_0$. For purpose of visualization we follow the degeneracy line of $a$ and $s$. We see that simultaneously the redshift of the transition and to a lesser degree the amplitude of the transition is modified. For $\zeta$ we note the expected behavior of only rescaling the overall amplitude of the $\Delta \alpha(z)/\alpha_0$. We do not show the impact of the offset $\kappa$ since the variation in terms of $\Delta \alpha/\alpha_0$ is very small.}
    \label{fig:alpha_zth}
\end{figure*}

\begin{figure*}[!h]
    \centering    \includegraphics[width=0.85\textwidth]{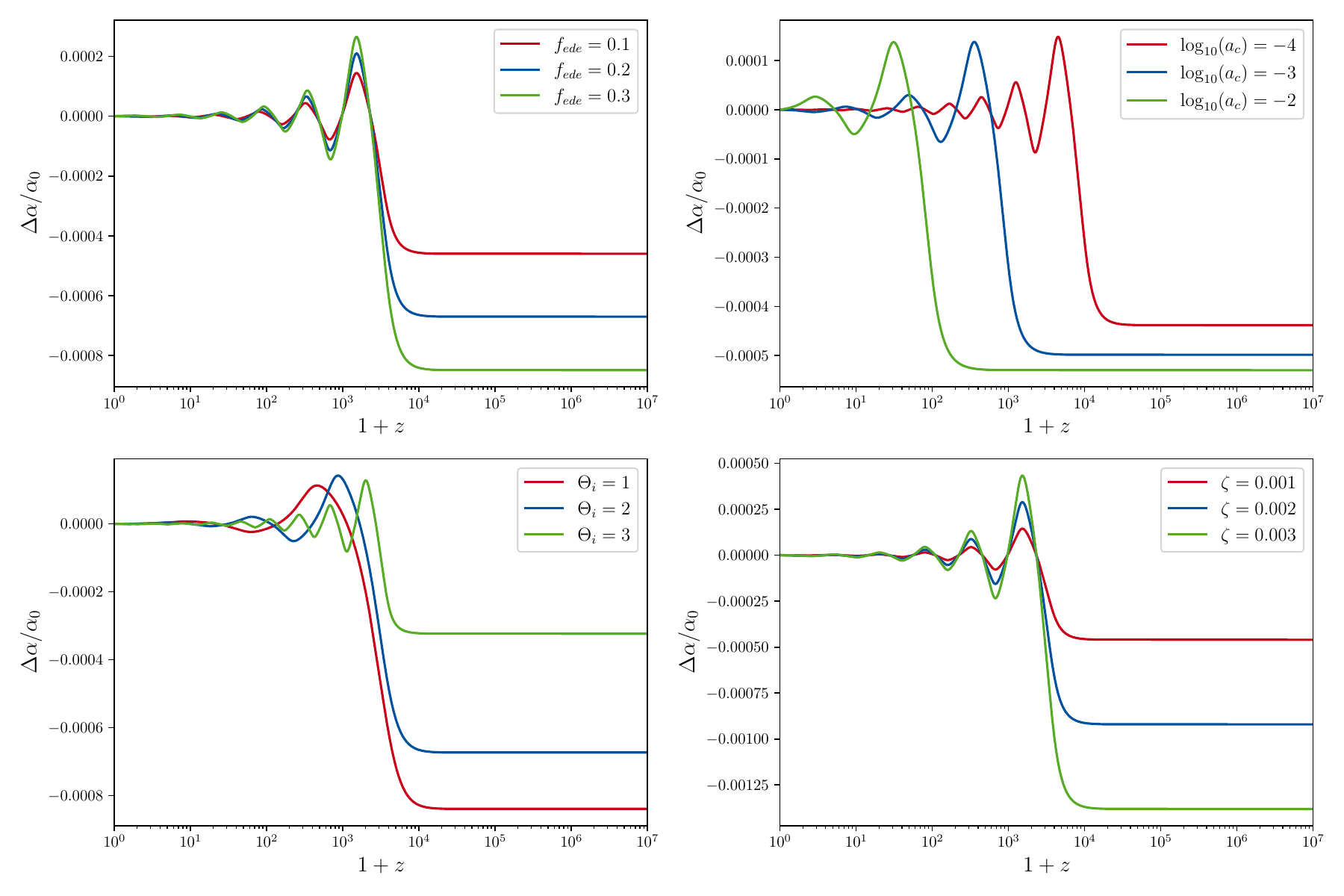}    
    \caption{Impact of the different parameters of the ALP model on the cosmic evolution of $\Delta \alpha(z)/\alpha_0$. We show only the additional model parameters, and note that the usual cosmological $\Lambda$CDM parameters do not impart large changes on $\Delta \alpha(z)/\alpha_0$\,. Both $f_\mathrm{ede}$ and $\zeta$ rescale the amplitude of the fine-structure constant variation, while $a_c = 1/(1+z_c)$ changes the redshift of the transition from constant (due to Hubble drag) to oscillating field. Finally, $\Theta_i$ changes the phase of the oscillations (as well as very slightly changing the initial redshift).}
    \label{fig:alpha_ALP}
\end{figure*}

In this appendix we display several examples of evolutions of $\Delta \alpha(z)/\alpha_0$ and their variation with the relevant underlying model parameters to facilitate the understanding of the main text. In \cref{fig:alpha_ALP} we show the evolutions for the ALP EDE model, while in \cref{fig:alpha_zth} we show the evolution for the $\tanh$ toy model of \cref{ssec:tanh_example}.


\end{document}